\documentclass{caosp306}

\usepackage{graphicx}

\usepackage{natbib}
\articleNo{}
\pubyear{}
\volume{}
\volnumber{}
\firstpage{1}
\received{}
\accepted{}

\def\BibTeX{{\rm B\kern-.05em{\sc i\kern-.025em b}\kern-.08em
             T\kern-.1667em\lower.7ex\hbox{E}\kern-.125emX}}

\begin{document}

%
\hauthor{T.\,{\.I}{\c c}li,D.\,Ko{\c c}ak and K.\,Yakut}

\title{Photometric study of selected X-ray binaries}


%
\author{
        Tu\u{g}\c{c}e {\.I}{\c c}li 
      \and 
         Dolunay Ko{\c c}ak 
      \and 
        Kadri Yakut 
       }
\institute{
           Department of Astronomy and Space Sciences, University of Ege, 35100, Bornova--{\.I}zmir, Turkey, \email{icli.tugce@gmail.com}
          }

\date{}

\maketitle

\begin{abstract}
We present results of a long-term photometric multicolor optical
monitoring project of selected low-mass and
high-mass X-ray binaries carried out at the T\"UB\.ITAK
National Observatory (TUG).  New long-term $VRI$ multicolor
observations of three selected X-ray binaries with neutron star
components (HZ Her, ScoX-1, SAX J2103.5+4545) were observed between
2015 and 2019 with the TUG 60-cm telescope. The light variations of the
systems are presented and discussed.

\end{abstract}

\section{Introduction}
\label{intr}

Flux variations of X-ray binaries consisting of an evolved component
and a compact object (neutron star/black hole) can be observed both at
X-ray and optical wavelengths. In these binary systems, the optical
component usually fills its Roche lobe. Long-term light variations of
both low-mass (LMXB) and high-mass (HMXB) X-ray binaries provide
information on astrophysical processes, particularly on hot stellar
winds of the companion star, stellar activity, mass transfer between
components, non-conservative mass loss and  angular momentum loss from the system. 
In this context, long-term multi-color optical changes of
selected low and high mass X-ray binaries with neutron star
components, previously discussed and catalogued by {\.I}\c{c}li \&
Yakut (2015), {\.I}\c{c}li (2016), are presented in this study.

Within the context of our observational project, eight X-ray binaries
with neutron star components (HZ Her, Sco X-1, PSR J1023+0038,
X Per, BQ Cam, V934 Her, SAX J2103.5+4545, XTE J1946+274) were
examined. Some parameters of the selected systems are summarized in
Table~1. Here we present long-term observational results for
HZ Her, Sco X-1, and SAX J2103.5 + 4545 and we obtain new period
for the SAX J2103.5+4545 system from our new observations.  
Sco X-1, which is known as the brightest LMXB, was discovered by Giacconi et al. (1962). Its optical component is V818 Sco (Sandage et al., 1966). 
The system has an orbital period of 0.78 days and consists of a 1.4 M$_{\odot}$
neutron star and a 0.42 M$_{\odot}$ optical component (Steeghs and
Casares, 2002). Based on optical observations, the spectral type is
thought to be earlier than G5.  The distance of the system is
2.8$\pm$0.3 kpc (Bradshaw et al., 1999) and the system exhibits a high
and low state (Bandyopadhyay et al., 1999).

Discovered by observations of the \texttt{UHURU} satellite in 1972
(Tananbaum et al., 1972), Her X-1 (HZ Her, 4U 1656+35) is classified
as an eclipsing LMXB system with an A7 spectral type (Middleditch \&
Nelson, 1976; Leahy \& Scott, 1998, {\.I}{\c{c}}li et al., 2019).  The system has an orbital period
of 1.7 days and consists of an accreting neutron star with a mass of
1.5 M$_{\odot}$ and an optical component with a mass of 2.2
M$_{\odot}$ (Reynolds et al., 1997; Leahy \& Abdallah, 2014, {\.I}{\c{c}}li et al., 2019).  Its
distance is 6.6 kpc (Reynolds et al., 1997). The system has been
observed at different wavelengths (optical, ultraviolet, radio and
X-ray bands), see, e.g., Shakura et al. (1997); Cherepashchuk et
al. (1974); Simon et al. (2002); {\.I}\c{c}li \& Yakut (2015);
{\.I}\c{c}li (2016). The X-ray flux exhibits a 35-day period variation related
to the accretion disc (Scott et al., 2000; Leahy \&
Abdallah, 2014; Postnov et al., 2013).  Another system within the scope of this
study is SAX J2103.5+4545, discovered in 1997 with BeppoSAX. It pulsates with a period of 358.61\,s (Hulleman et
al.,1998). The system is composed of a 20 M$_{\odot}$ optical component
with a B0\,Ve spectral type (Reig et al., 2010). The orbital period of the
binary system is 12.6 days (Baykal et al., 2007).

\section{New observations}
New observations were made between February 2015 and July
2019, with varying exposure times between 5, 20 and 60 seconds.  
Multicolour ($VRI$) light curves were obtained with the 60-cm robotic telescope at the T\"UB\.ITAK National Observatory (TUG). Three systems were observed, HZ Her, ScoX-1, and
SAX J2103.5+4545. Their respective periods are 1.7, 0.78, 12.6 days, and their $V$
band brightness lies between 12.5 and 13.8 magnitudes.
Observations were handled with the standard difference photometry method in each
observation term. IRAF/PHOT and AstroImageJ (Collins et al., 2017)
were used in the reduction stage. The frame reduction was performed by
subtracting the bias and dark frames and dividing by
flat-field frames. Following the time correction, we performed differential photometry as we did in our prevous studies ({\.I}{\c{c}}li et al., 2013, {\c{C}}okluk, et al., 2019, Ko{\c{c}}ak et al. 2019). AAVSO-135, AAVSO-132, AAVSO-50(1) (for HZ Her),
AAVSO-115, AAVSO-126, AAVSO-113 (for ScoX-1) and ID-72,131,136 (for
SAX J2103.5+4545 (Reig \& Fabregat, 2015) were chosen as comparison
stars. Light variations of the systems are plotted in
Fig.~\ref{figures}.


\begin{table}[t]
\centering
\setlength{\tabcolsep}{0.025in}
\begin{center}
\caption{Parameters of selected X-ray binaries}
\label{t1}
\scriptsize
\begin{tabular}{llcccclcl}
\hline
System          &Alias    &Type &{$\alpha$} &{$\delta$}   	   & $V$ (mag)&$P_{\rm orb}$ (d)  &OTime & N$_{\rm obs}$\\
\hline
PSR J1023+0038   &AY Sex   &LMXB &10 23 48    & +00 38 41       &17.5 &0.198&7731-8447&337 \\
Sco X-1         &V818 Sco &LMXB &16 19 43    & $-$15 39 08  	   &12.5 &0.787&7106-8644&847 \\
Her X-1         &HZ Her   &LMXB &16 57 49    & +35 20 33       &13.6 &1.7  &7236-8687&1145\\
SAX J2103.5+4545 &         &HMXB &21 02 55    & +45 43 25       &14.2 &12.6 &7236-8461&1029\\
V0332+52        &BQ Cam   &HMXB &03 34 60    & +53 10 23       &15.4 &34.67&7059-8759&2126\\
XTE J1946+274    &         &HMXB &19 45 35    & +27 20 43       &16.9 &169.2&7271-8659&466 \\
3A 0352+309     &X Per    &HMXB &03 55 36    & +31 00 25       &6.7  &250.3&7360-8461&620 \\
4U 1700+24       &V934 Her &LMXB &17 06 35    & +23 58 19       &7.6  &4391 &7137-8461&1230\\
\hline
\end{tabular}
\end{center}
\end{table}

\section{Discussion and conclusions}

\begin{figure}
\includegraphics[width=0.99\textwidth,clip=]{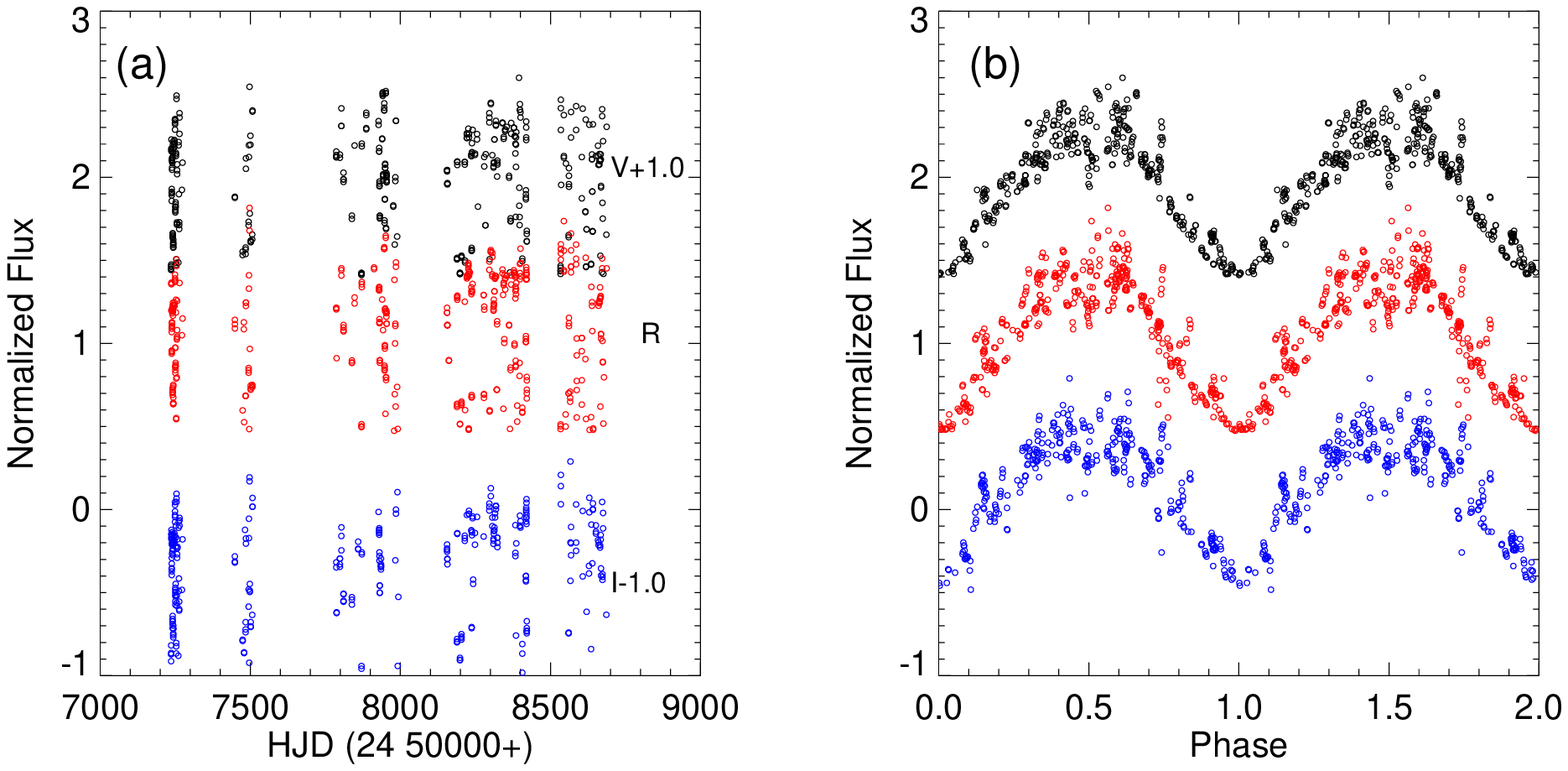}\\
\includegraphics[width=0.99\textwidth,clip=]{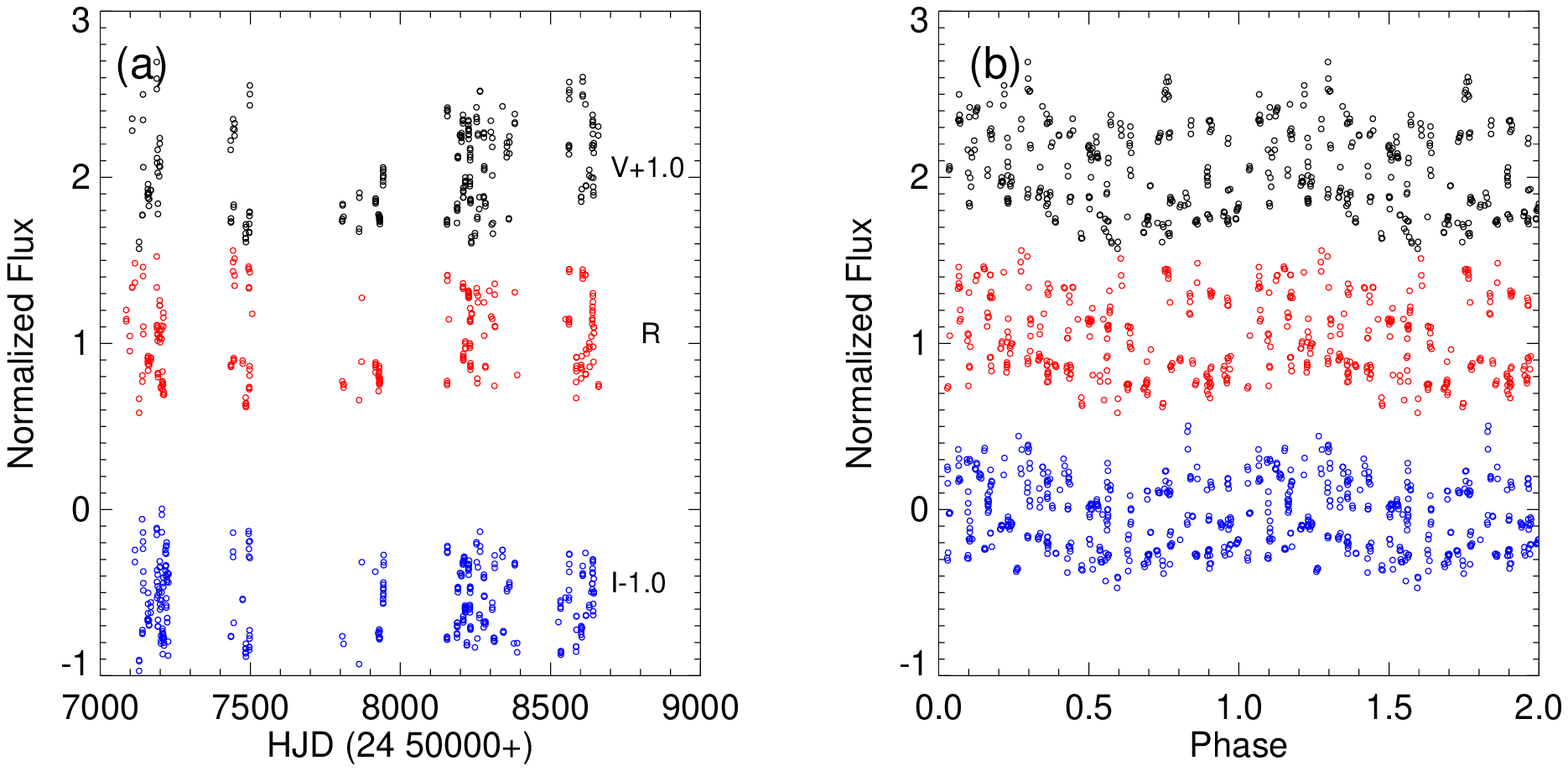}\\
\includegraphics[width=0.99\textwidth,clip=]{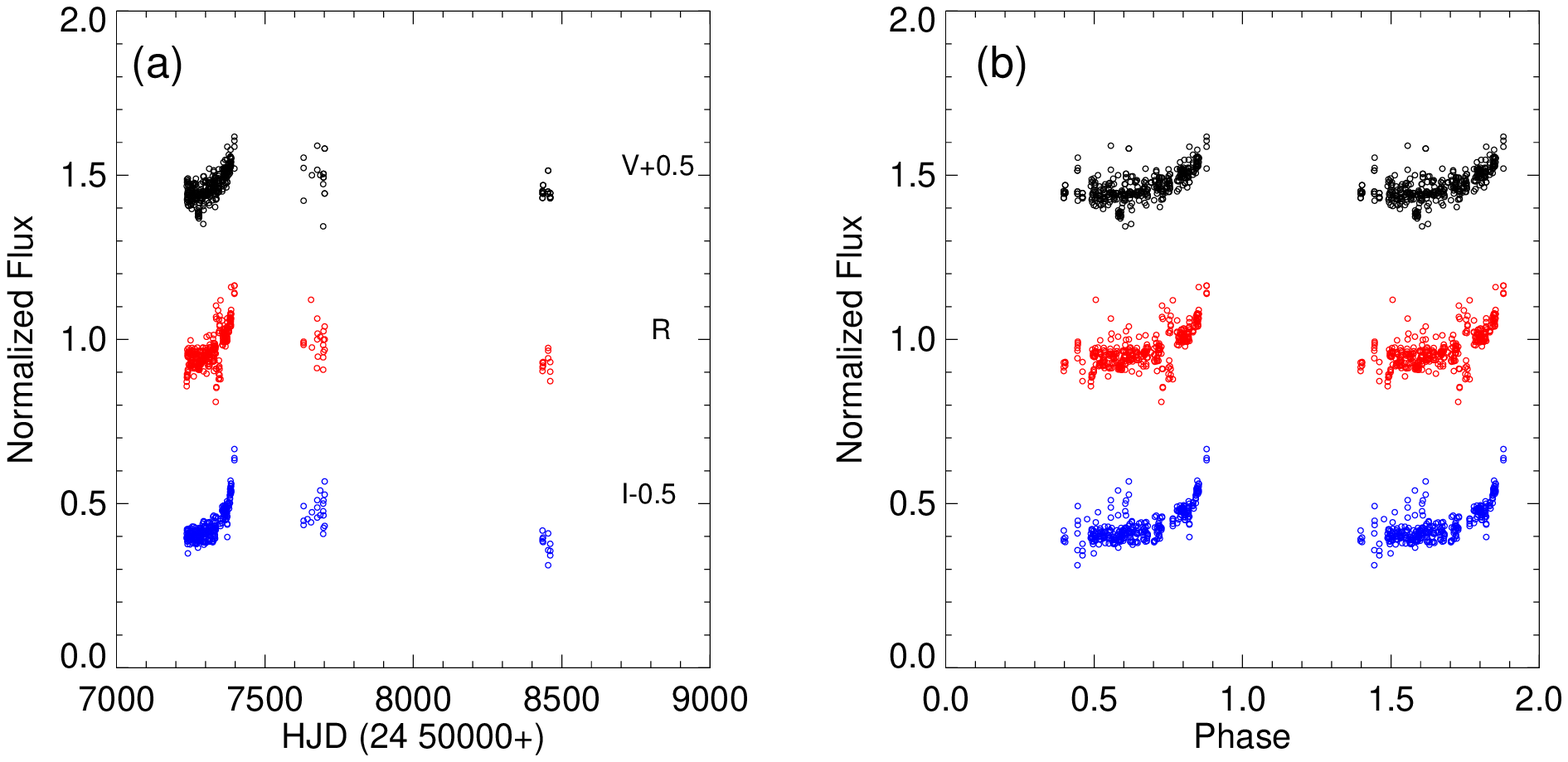}
\caption{Long-term $VRI$ light variations of the systems in the time (a) and
  phase (b) domain for HZ Her (top), Sco X-1 (middle), and
  SAX J2103.5+4545 (bottom).}
\label{figures}
\end{figure}

	X-ray binary systems consist of an early- or  a late-type star and a black hole or neutron star. 
	The presence of an accretion disk, mass loss from the system and stellar activity on the companion star affect the multi-wavelength light curves of the binary systems.
	We obtained multicolor ($VRI$) observations of X-ray binaries with neutron star components between 2015 and 2019 with the TUG T60 telescope. Periodic and non-periodic changes have been observed in these systems.  
	For HZ Her, the amplitude of the light variation in the $V$ band is 1.5 mag.  The middle panel of Figure~\ref{figures} shows the multicolor optical light variation of V818 Sco. The light curves of the system show amplitude variability in the VRI bands. Orbital light curves of the V818 Sco (Figure~\ref{figures} middle right panel) can differ significantly from the mean orbital light variation ($\sim 35\%$). 
	Based on observations obtained over a span of five years, the period of the HMXB SAX J2103.5+4545 was found to be variable. To determine new periods of the system, we used Period04 (Lenz \& Breger 2005) software. A new non-periodic variation was determined to be 412 days from the new long-term observations.


\acknowledgements 

We are very grateful to an anonymous referee for comments and suggestions. This study is supported by the Turkish Scientific
and Research Council (T\"UB\.ITAK - 117F188).  We thank T\"UB\.ITAK
for partial support in using the T60 telescope with project numbers
15AT60-776 and 18AT160-1298. T\.I thanks T\"{U}B\.{I}TAK for his Fellowship (2211-C). KY would like to acknowledge the contribution of  COST (European Cooperation in Science and Technology) Action CA15117 and CA16104.


\begin{thebibliography}{}

\bibitem[Bandyopadhyay et al.(1999)]{1999MNRAS.306..417B} Bandyopadhyay, R.~M., Shahbaz, T., Charles, P.~A., et al.\ 1999, \mnras, 306, 417
\bibitem[Baykal et al.(2007)]{2007MNRAS.374.1108B} Baykal, A., Inam, S. {\c{C}}., Stark, M.~J., et al.\ 2007, \mnras, 374, 1108
\bibitem[Bradshaw et al.(1999)]{1999ApJ...512L.121B} Bradshaw, C.~F., Fomalont, E.~B., \& Geldzahler, B.~J.\ 1999, \apjl, 512, L121
\bibitem[Cherepashchuk et al.(1974)]{1974PZ.....19..305C} Cherepashchuk, A.~M., Kovalenko, V.~M., Kovalenko, O.~N., et al.\ 1974, Peremennye Zvezdy, 19, 305
\bibitem[Collins et al.(2017)]{2017AJ....153...77C} Collins, K.~A., Kielkopf, J.~F., Stassun, K.~G., et al.\ 2017, \aj, 153, 77
\bibitem[\protect\citeauthoryear{{\c{C}}okluk, et al.}{2019}]{2019MNRAS.488.4520C} {\c{C}}okluk K.~A., Ko{\c{c}}ak D., I{\c{c}}li T., Karak{\"o}se S., {\"U}st{\"u}nda{\v{g}} S., Yakut K., 2019, MNRAS, 488, 4520
\bibitem[Giacconi et al.(1962)]{1962PhRvL...9..439G} Giacconi, R., Gursky, H., Paolini, F.~R., et al.\ 1962, \prl, 9, 439
\bibitem[Hulleman et al.(1998)]{1998A&A...337L..25H} Hulleman, F., in 't Zand, J.~J.~M., \& Heise, J.\ 1998, \aap, 337, L25
\bibitem[\protect\citeauthoryear{Icli}{2016}]{} \.I\c{c}li, T., 2016, MSc Thesis, Binary systems with neutron star components, University of Ege
\bibitem[Icli, \& Yakut(2015)]{2015ebha.confE..89I} \.I\c{c}li, T., \& Yakut, K.\ 2015, The Extremes of Black Hole Accretion, 89
\bibitem[\protect\citeauthoryear{{\.I}{\c{c}}li, Ko{\c{c}}ak \& Yakut}{2019}]{2019IAUS..346..239I} {\.I}{\c{c}}li T., Ko{\c{c}}ak D., Yakut K., 2019, IAUS,  239, IAUS..346
\bibitem[\protect\citeauthoryear{{\.I}{\c{c}}li, et al.}{2013}]{2013AJ....145..127I} {\.I}{\c{c}}li T., Ko{\c{c}}ak D., Boz G. {\c{C}}., Yakut K., 2013, AJ, 145, 127
\bibitem[\protect\citeauthoryear{Ko{\c{c}}ak, {\.I}{\c{c}}li \& Yakut}{2019}]{2019IAUS..346..252K} Ko{\c{c}}ak D., {\.I}{\c{c}}li T., Yakut K., 2019, IAUS,  252, IAUS..346
\bibitem[Leahy, \& Abdallah(2014)]{2014ApJ...793...79L} Leahy, D.~A., \& Abdallah, M.~H.\ 2014, \apj, 793, 79
\bibitem[Leahy, \& Scott(1998)]{1998ApJ...503L..63L} Leahy, D.~A., \& Scott, D.~M.\ 1998, \apjl, 503, L63
\bibitem[Middleditch, \& Nelson(1976)]{1976ApJ...208..567M} Middleditch, J., \& Nelson, J.\ 1976, \apj, 208, 567
\bibitem[Postnov et al.(2013)]{2013MNRAS.435.1147P} Postnov, K., Shakura, N., Staubert, R., et al.\ 2013, \mnras, 435, 1147
\bibitem[Reig et al.(2010)]{2010MNRAS.401...55R} Reig, P., S{\l}owikowska, A., Zezas, A., et al.\ 2010, \mnras, 401, 55
\bibitem[Reig, \& Fabregat(2015)]{2015A&A...574A..33R} Reig, P., \& Fabregat, J.\ 2015, \aap, 574, A33
\bibitem[Reynolds et al.(1997)]{1997MNRAS.288...43R} Reynolds, A.~P., Quaintrell, H., Still, M.~D., et al.\ 1997, \mnras, 288, 43
\bibitem[Sandage et al.(1966)]{1966ApJ...146..316S} Sandage, A., Osmer, P., Giacconi, R., et al.\ 1966, \apj, 146, 316
\bibitem[Scott et al.(2000)]{2000ApJ...539..392S} Scott, D.~M., Leahy, D.~A., \& Wilson, R.~B.\ 2000, \apj, 539, 392
\bibitem[Shakura et al.(1997)]{1997ASPC..121..379S} Shakura, N.~I., Smirnov, A.~V., \& Ketsaris, N.~A.\ 1997, IAU Colloq. 163: Accretion Phenomena and Related Outflows, 379
\bibitem[{\v{S}}imon et al.(2002)]{2002NewA....7..349S} {\v{S}}imon, V., Kroll, P., Neugebauer, P., et al.\ 2002, NewA, 7, 349
\bibitem[Steeghs, \& Casares(2002)]{2002ApJ...568..273S} Steeghs, D., \& Casares, J.\ 2002, \apj, 568, 273
\bibitem[Tananbaum et al.(1972)]{1972ApJ...174L.143T} Tananbaum, H., Gursky, H., Kellogg, E.~M., et al.\ 1972, \apjl, 174, L143

\end{thebibliography}

\end{document}